\newfam\msbfam
\font\twlmsb=msbm10 at 12pt
\font\eightmsb=msbm10 at 8pt
\font\sixmsb=msbm10 at 6pt
\textfont\msbfam=\twlmsb
\scriptfont\msbfam=\eightmsb
\scriptscriptfont\msbfam=\sixmsb

\centerline{\bf Locally inertial coordinates with totally antisymmetric torsion} 

\

\

\centerline{M. Socolovsky*}

\

\centerline{\it  Instituto de Ciencias Nucleares, Universidad Nacional Aut\'onoma de M\'exico}
\centerline{\it Circuito Exterior, Ciudad Universitaria, 04510, M\'exico D. F., M\'exico} 

\

{\bf Abstract} {\it We show that the necessary and sufficient condition for erecting locally inertial coordinates at a point $p$ of a $U^4$-space, and therefore assuring the validity of the equivalence principle at that point, is the vanishing at $p$ of the symmetric part of the contortion tensor. This fact does not demand a vanishing torsion, but only a totally antisymmetric one. As an application, we derive the geodesic deviation equation; and prove the compatibility with the Newtonian limit.}

\


\

{\bf 1. Introduction}

\

The problem of the validity of the equivalence principle (EP) in gravity theories with torsion and zero non-metricity [1], [2], has been discussed by several authors [3], [4], [5]. The strategy of these authors consisted in showing the existence of local basis or normal frames such that, with respect to them, all components of the linear connection, which gives the gravitational ``force", vanish at a given point of the manifold (von der Heyde, Hartley) or also in a neighborhood of the point (Iliev). The purpose of the present article is to show that a particularly simple change of coordinates, can also maintain the validity of the EP in the presence of torsion, with the only restriction of this being totally antisymmetric. 

\

This is allowed by the structure of the geodesic equation. Using the fact that only the symmetric part of the connection intervenes in this equation, in section {\bf 2} we show that a necessary and sufficient condition for the vanishing of the first derivatives of the metric is the vanishing of the symmetric part of the contortion tensor, but not of the torsion tensor itself, which is only required to be totally antisymmetric. Then, in the transformed coordinates, the Levi-Civita part of the connection also vanishes and the geodesic equation corresponds to a freely falling system. 

\

The geodesic deviation equation is an instrument to detect the presence of curvature and/or torsion in the manifold (spacetime). In particular, it is relevant for the studies of gravitational waves detection [6], [7]. In section {\bf 3} we derive this equation in a Riemann-Cartan space ($U^n$-space); in particular, we specialize it for the case of a totally antisymmetric torsion and $n$=4. 

\

Finally, in section {\bf 4}, we show the compatibility between a totally antisymmetric torsion and the Newtonian limit of the geodesic equation. 

\

Section {\bf 5} is devoted to additional comments.

\

{\bf 2. Equivalence principle}
 
\

Let $(M^n,g,\Gamma)$ be a $U^n$-space i.e. a real differentiable manifold $M^n$ with metric $g$ and connection $\Gamma$ compatible with $g$, $p\in M^n$, and $(U,\varphi=(x^\mu))$ a chart on $M^n$ with $p\in U$ and $x^\mu(p)=0$, $\mu=0,\dots,n-1$. Let $(U^\prime,\varphi^\prime=(x^{\prime\mu}))$ be an intersecting chart with $x^{\prime\mu}(p)=0$ and [8] $$x^\mu=x^{\prime\mu}-{{1}\over{2}}\Gamma^\mu_{(\nu\rho)}(p)x^{\prime\nu}x^{\prime\rho},\eqno{(1)}$$ where $(\nu\rho)$ means symmetrization. The antisymmetric part of the connection, $\Gamma^\mu_{[\nu\rho]}=T^\mu_{\nu\rho}=-T^\mu_{\rho\nu}$ : torsion, does not contribute to the above change of coordinates. (Units: $[\Gamma^\mu_{\nu\rho}]=[T^\mu_{\nu\rho}]=[L]^{-1}$.) 

\

The condition of metricity at $p$: $$0=g_{\mu\nu ;\lambda}(p)=g_{\mu\nu ,\lambda}(p)-\Gamma^\rho_{\lambda\mu}(p)g_{\nu\rho}(p)-\Gamma^\rho_{\lambda\nu}(p)g_{\mu\rho}(p),\eqno{(2)}$$ which being a tensor also holds in $U \cap U^\prime$, the tensor transformation formula of $g_{\mu\nu}$, and the diagonalization of $g_{\mu\nu}$ to $\eta_{\mu\nu}$ at $p$, lead to the equations: $$g^{\prime}_{\mu\nu}(x^{\prime\lambda})=\eta_{\mu\nu}+(\eta_{\mu\rho}T^\rho_{\lambda\nu}(p)+\eta_{\nu\rho}T^\rho_{\lambda\mu}(p))x^{\prime\lambda}+O({x^{\prime\mu}}^2)=\eta_{\mu\nu}+{{\partial}\over{\partial x^{\prime\lambda}}}g^\prime_{\mu\nu}(p)x^{\prime\lambda}+O({x^{\prime\mu}}^2), \eqno{(3)}$$ and $${(\Gamma_{LC}^\prime)}^\mu_{\nu\rho}(p)={{1}\over{2}}\eta^{\mu\sigma}(\partial^\prime_\nu (g^\prime_{\rho\sigma})(p)+\partial^\prime_\rho (g^\prime_{\sigma\nu})(p)-\partial^\prime_\sigma (g^\prime_{\nu\rho})(p)). \eqno{(4)}$$ So, $$T^\mu_{\nu\rho}(p)=0 \Rightarrow \partial^\prime_\lambda (g^\prime_{\mu\nu})(p)=0  \  \ and \ \ 
then \ \ g^\prime_{\mu\nu}(x^{\prime\lambda})=\eta_{\mu\nu}+O({x^{\prime\mu}}^2) \ \ and \ \ {(\Gamma^\prime_{LC})}^\mu_{\nu\rho}(p)=0,\eqno{(5)}$$ i.e. the vanishing of the torsion at $p$ is a {\it sufficient} condition for having a local inertial system at $p$. 

\ 

However, the condition is not {\it necessary}: in fact, $$\eta_{\mu\rho}T^\rho_{\lambda\nu}(p)+\eta_{\nu\rho}T^\rho_{\lambda\mu}(p)=T_{\lambda\nu\mu}(p)+T_{\lambda\mu\nu}(p)=0 \eqno{(6)}$$ implies that $T_{\mu\nu\rho}$ is also antisymmetric in its second and third indices, and then it is totally antisymmetric, since $T_{\mu\nu\lambda}=-T_{\mu\lambda\nu}=T_{\lambda\mu\nu}=-T_{\lambda\nu\mu}$. 

\

A calculation gives: 

\

$n=2$:$$T^0_{01}=T^1_{01}=0 \eqno{(7)}$$

\ 

$n=3$: $$T^0_{01}=T^0_{02}=T^1_{01}=T^1_{12}=T^2_{02}=T^2_{12}=0,$$ $$T^0_{12}=T^2_{10}=T^1_{02} \eqno{(8)}$$

\

$n=4$: $$T^0_{01}=T^0_{02}=T^0_{03}=T^1_{01}=T^1_{12}=T^1_{13}=T^2_{02}=T^2_{12}=T^2_{32}=T^3_{03}=T^3_{13}=T^3_{23}=0,$$ $$T^0_{12}=T^2_{10}=T^1_{02},$$ $$T^0_{13}=T^3_{10}=T^1_{03},$$ $$T^0_{23}=T^3_{20}=T^2_{03},$$ $$T^1_{23}=T^2_{31}=T^3_{12} \eqno{(9)}$$ In each case, the number of independent but not necessarily zero components of the torsion tensor coincides with the number of independent components of the totally antisymmetric torsion tensor with covariant indices, number which results from the condition that geodesics defined as ``world-lines of particles" (parallel transported velocities) coincide with their definition as extremals of arc-length [2]. This last fact can be seen as follows:

\ 

As world-lines, geodesics are defined as $${{d^2 x^\alpha}\over{d\lambda ^2}}+\Gamma^\alpha_{(\nu\mu)}{{dx^\nu}\over{d\lambda}}{{dx^\mu}\over{d\lambda}}=0 \eqno{(10)}$$ where only the symmetric part of $\Gamma^\mu_{\nu\rho}$ contributes: $$\Gamma^\alpha_{(\nu\mu)}={(\Gamma_{LC})}^\alpha_{\nu\mu}+K^\alpha_{(\nu\mu)}=
{(\Gamma_{LC})}^\alpha_{\nu\mu}-g^{\alpha\rho}(T^\lambda_{\mu\rho}g_{\lambda\nu}+T^\lambda_{\nu\rho}g_{\lambda\mu})={(\Gamma_{LC})}^\alpha_{\nu\mu}-(T_\mu \ ^\alpha \ _\nu +T_\nu \ ^\alpha \ _\mu ) \eqno{(11)}$$ with $g^{\delta\gamma}T^\alpha_{\beta\gamma}g_{\alpha\sigma}=g^{\delta\gamma}T_{\beta\gamma\sigma}=T_\beta \ ^\delta \ _\sigma$ and $K^\alpha_{\nu\mu}=T^\alpha_{\nu\mu}+K^\alpha_{(\nu\mu)}$ the contortion tensor; notice that the covariant form of the torsion tensor, $T_{\beta\gamma\delta}$, is antisymmetric in the first two indices: $T_{\beta\gamma\delta}=-T_{\gamma\beta\delta}$. With this definition of $T_{\alpha\beta\gamma}$, the covariant form of the contortion tensor is $$K_{\mu\nu\rho}=g_{\rho\alpha}K^\alpha_{\mu\nu}=T_{\mu\nu\rho}-T_{\nu\rho\mu}+T_{\rho\mu\nu}, \eqno{(12)}$$ which is antisymmetric in the last two indices i.e. $K_{\mu\nu\rho}=-K_{\mu\rho\nu}$.

\

On the other hand, the equation of geodesics defined as extremals of arc-length: $$0=\delta\int ds=\delta\int(g_{\mu\nu}dx^\mu dx^\nu )^{{1}\over{2}},\eqno{(13)}$$ turns out to be [9] $${{d^2 x^\alpha}\over{d\lambda ^2}}+{(\Gamma_{LC})}^\alpha_{\nu\mu}{{dx^\nu}\over{d\lambda}}{{dx^\mu}\over{d\lambda}}=0. \eqno{(14)}$$ Then, for the definitions $(10)$ and $(14)$ to coincide, $T_{(\mu} \ ^\alpha \ _{\nu )}$ must vanish i.e. $T_\mu \ ^\alpha \ _\nu =-T_\nu \ ^\alpha \ _\mu \Leftrightarrow g^{\alpha\rho}T_{\mu\rho\nu}=-g^{\alpha\rho}T_{\nu\rho\mu} \Leftrightarrow T_{\mu\sigma\nu}=-T_{\nu\sigma\mu}$ i.e. $T_{\alpha\beta\gamma}$ must be 1-3 antisymmetric; but this implies that $T_{\alpha\beta\gamma}$ is also 2-3 antisymmetric: $T_{\mu\sigma\nu}=-T_{\nu\sigma\mu}=T_{\sigma\nu\mu}=-T_{\mu\nu\sigma}$. Since, by definition, $T_{\mu\nu\rho}$ is antisymmetric in the first two indices, it turns out that $T_{\mu\nu\rho}$ is totally antisymmetric; in $n$ dimensions, its number of independent components is $\pmatrix{n \cr 3 \cr}={{n(n-1)(n-2)}\over{6}}\equiv N$. Some values are: $$\matrix{n & 2 & 3 & 4 & 5 & \cdots \cr N & 0 & 1 & 4 & 10 & \cdots \cr}$$ As independent components we can choose, for $n=4$, $T_{120}$, $T_{230}$, $T_{310}$ and $T_{231}$.

\

The set of allowed non-vanishing components of the torsion tensor still leads to ``physical" (geometrical) effects. The non-closure of a parallelogram with infinitesimal sides $\epsilon^\mu$ and $\delta^\nu$ is measured by the vector $$\Delta^\mu=2T^\mu_{\beta\alpha}\delta^\beta\epsilon^\alpha=T^\mu_{\beta\alpha}(\delta^\beta\epsilon^\alpha-\delta^\alpha\epsilon^\beta).\eqno{(15)}$$ In particular, for $n=4$, its components are:

$$\Delta^0=T^0_{12}(\delta^1\epsilon^2-\delta^2\epsilon^1)+T^0_{13}(\delta^1\epsilon^3-\delta^3\epsilon^1)+T^0_{23}(\delta^2\epsilon^3-\delta^3\epsilon^2),$$
$$\Delta^1=T^1_{23}(\delta^2\epsilon^3-\delta^3\epsilon^2)+T^1_{02}(\delta^0\epsilon^2-\delta^2\epsilon^0)+T^1_{03}(\delta^0\epsilon^3-\delta^3\epsilon^0),$$
$$\Delta^2=T^2_{31}(\delta^3\epsilon^1-\delta^1\epsilon^3)+T^2_{10}(\delta^1\epsilon^0-\delta^0\epsilon^1)+T^2_{03}(\delta^0\epsilon^3-\delta^3\epsilon^0),$$
$$\Delta^3=T^3_{10}(\delta^1\epsilon^0-\delta^0\epsilon^1)+T^3_{20}(\delta^2\epsilon^0-\delta^0\epsilon^2)+T^3_{12}(\delta^1\epsilon^2-\delta^2\epsilon^1),\eqno{(16)}$$ which can be distinct from zero. For example, in arbitrary units, let $\epsilon^\alpha=(1,1,2,0)$ and $\delta^\beta=(1,2,0,0)$; then $\Delta^\mu=(4\Delta,2\Delta,\Delta,T^3_{10}-2T^3_{20}+4T^3_{12})$, with $\Delta=T^0_{12}$. 

\

In summary, the {\it necessary} and {\it sufficient} condition for erecting a locally inertial {\it coordinate} system at a point $p$ in a $U^4$-space, is that the symmetric part of the contortion tensor vanish at $p$, i.e. $$K^\alpha_{(\mu\nu)}(p)=\eta^{\alpha\rho}(T^\lambda_{\rho\mu}(p)\eta_{\lambda\nu}+T^\lambda_{\rho\nu}(p)\eta_{\lambda\mu})=0.\eqno{(17)}$$ 

\

Therefore, the {\it equivalence principle} still {\it holds} in a $U^4$-space ($U^4=(M^4,g,\Gamma))$ with $\Gamma$ a metric connection with totally antisymmetric torsion tensor.

\

{\bf 3. Geodesic deviation}

\

Once we allow for a totally antisymmetric torsion tensor, it is interesting to see how it modifies the geodesic deviation equation (GDE) with the aim, in principle, of measuring torsion during the free fall of nearby objects [10]. The GDE is an equation which gives the relative acceleration between nearby geodesics in terms of curvature (in a $V^n$-theory), of torsion (in a $T^n$-theory), or of curvature and torsion (in a $U^n$-theory). 

\

Let $x^\mu (s,t)$ be a family of geodesics with affine parameter $t$ and indexed by $s$. $s$ and $t$ are local coordinates of a two-dimensional surface embedded in the manifold $M^n$; from now on, we shall only consider the case $n=4$. The vectors (fields) $${{\partial} \over {\partial t}}={{\partial x^\mu} \over {\partial t}} \partial _\mu =T^\mu \partial _\mu =T \eqno{(18)}$$ and $${{\partial}\over {\partial s}}={{\partial x^\mu} \over {\partial s}}\partial_\mu=S^\mu \partial_\mu=S \eqno{(19)}$$ are respectively the {\it tangent vector} to the geodesic $s$ and the {\it deviation vector} of the same geodesic at the affine parameter value $t$. ($[T^\mu]=[L]^0$, $[S^\mu]=[L]$.) Since $\partial_t$ and $\partial_s$ commute, then $$0=[S,T]=[S,T]^\nu\partial_\nu \eqno{(20)}$$ with $$[S,T]^\nu=S^\mu {T^\nu}_{,\mu}-T^\mu{S^\nu}_{,\mu}=S^\mu ({T^\nu}_{;\mu}-\Gamma^\nu_{\mu\rho}T^\rho)-T^\mu({S^\nu}_{;\mu}-\Gamma^\nu_{\mu\rho}S^\rho)=S^\mu{T^\nu}_{;\mu}-T^\mu{S^\nu}_{;\mu}-2S^\mu T^\nu_{\mu\rho}T^\rho \eqno{(21)}$$ and therefore $$S^\mu{T^\nu}_{;\mu}=T^\mu{S^\nu}_{;\mu}+2S^\mu T^\nu_{\mu\rho}T^\rho. \eqno{(22)}$$ The {\it relative velocity} and {\it relative acceleration} between two nearby geodesics are respectively defined by $$V=\nabla_T S=V^\mu\partial_\mu, \ \ V^\mu=T^\nu{S^\mu}_{;\nu}, \ \ \ ([V^\rho]=[L]^0),\eqno{(23)}$$ and $$A=\nabla_T V=A^\mu\partial_\mu ,  \ \ \ ([A^\mu]=[L]^{-1}),\eqno{(24)}$$ with $$A^\mu=T^\nu{V^\mu}_{;\nu}=T^\nu (T^\rho{S^\mu}_{;\rho})_{;\nu}=(T^\rho{S^\sigma}_{;\rho}){T^\mu}_{;\sigma}+T^\rho S^\sigma ({T^\mu}_{;\sigma})_{;\rho}$$ $$-2(T^\rho {S^\sigma}_{;\rho}T^\mu_{\sigma\lambda}T^\lambda+T^\rho S^\sigma{T^\mu_{\sigma\lambda}}_{;\rho}T^\lambda+T^\rho S^\sigma T^\mu_{\sigma\lambda}{T^\lambda}_{;\rho}).\eqno{(25)}$$ Using the commutator of covariant derivatives $$[D_\rho,D_\sigma]T^\mu\equiv ({T^\mu}_{;[\sigma})_{;\rho]}=R^\mu_{\lambda\rho\sigma}T^\lambda-2T^\lambda_{\rho\sigma}{T^\mu}_{;\lambda} \eqno{(26)}$$ where the curvature tensor is given by $$R^\rho_{\sigma\mu\nu}=\Gamma^\rho_{[\nu \sigma, \mu]}+\Gamma^\rho_{[\mu \lambda}\Gamma^\lambda_{\nu] \sigma}, \ \ \ ([R^\rho_{\sigma\mu\nu}]=[L]^{-2}),\eqno{(27)}$$ the Leibnitz rule, equation $(22)$, and the fact that for geodesics $T^\rho{T^\lambda}_{;\rho}=0$, one finds the acceleration $$A^\mu =A^\mu _R +A^\mu _T,\eqno{(28)}$$ with $$A^\mu _R=R^\mu_{\lambda\rho\sigma}T^\lambda T^\rho S^\sigma,\eqno{(29)}$$ the {\it curvature} part, and $$A^\mu _T=2T^\lambda (S^\sigma T^\mu_{\lambda\sigma})_{;\rho}T^\rho,\eqno{(30)}$$ the {\it torsion} part. For a metric connection, we can write $$A^\mu_T=2g^{\mu\nu}T^\lambda{(S^\sigma T_{\lambda\sigma\nu})}_{;\rho}T^\rho.\eqno{(31)}$$ Using $T_{120}$, $T_{230}$, $T_{310}$ and $T_{231}$ as the independent components of the totally antisymmetric torsion tensor $T_{\mu\nu\rho}$, a straightforward calculation leads to the result: $$\pmatrix{A^0_T \cr A^1_T \cr A^2_T \cr A^3_T}=2T^\rho\pmatrix{g^{00} & g^{01} & g^{02} & g^{03} \cr g^{01} & g^{11} & g^{12} & g^{13} \cr g^{02} & g^{12} & g^{22} & g^{23} \cr g^{03} & g^{13} & g^{23} & g^{33}}\pmatrix{{\cal A}_\rho \cr {\cal B}_\rho \cr {\cal C}_\rho \cr {\cal D}_\rho},\eqno{(32)}$$ with $${\cal A}_\rho=T^{[1}(S^{2]}T_{120})_{;\rho}+T^{[2}(S^{3]}T_{230})_{;\rho}+T^{[3}(S^{1]}T_{310})_{;\rho},$$ $${\cal B}_\rho=T^{[0}(S^{3]}T_{310})_{;\rho}+T^{[2}(S^{0]}T_{120})_{;\rho}+T^{[2}(S^{3]}T_{231})_{;\rho},$$ $${\cal C}_\rho=T^{[0}(S^{1]}T_{120})_{;\rho}+T^{[3}(S^{0]}T_{230})_{;\rho}+T^{[3}(S^{1]}T_{231})_{;\rho},$$ $${\cal D}_\rho=T^{[0}(S^{2]}T_{230})_{;\rho}+T^{[1}(S^{0]}T_{310})_{;\rho}+T^{[1}(S^{2]}T_{231})_{;\rho}.\eqno{(33)}$$ 

\

{\bf 4. Newtonian limit}

\

In this section we show that the Newtonian limit of the geodesic equation requires the vanishing of the components $T^0_{01}$, $T^0_{02}$ and $T^0_{03}$ of the torsion tensor; this fact is compatible with the totally antisymmetric torsion of section {\bf 2}.

\

Consider a timelike geodesic with affine parameter the proper time $\tau$. For low particle velocities, $\vert{{dx^i}\over{d\tau}}\vert <<\vert{{dt}\over{d\tau}}\vert$ and the equation becomes $${{d^2 x^\mu}\over{d\tau ^2}}+\Gamma^\mu_{00}({{dt}\over{d\tau}})^2=0 \ \ or \ \ \ \ddot{x}^\mu(t)+{(\Gamma_{LC})}^\mu_{00}+K^\mu_{00}=0 \eqno{(34)}$$ with $${(\Gamma_{LC})}^\mu_{00}=-{{1}\over{2}}g^{\mu\lambda}\partial_\lambda g_{00} \ \ \ and \ \ \ K^\mu_{00}=2g^{\mu\lambda}T^\sigma_{\lambda 0}g_{\sigma 0}.\eqno{(35)}$$ For small gravitational fields $g_{\mu\nu}=\eta_{\mu\nu}+h_{\mu\nu}$ with $\vert h_{\mu\nu}\vert <<1$, and for static solutions $g_{k0}=g^{k0}=0$ (to guarantee time-reversal invariance) and ${{\partial}\over {\partial t}}h_{\mu\nu}=0$. Then $${(\Gamma_{LC})}^0_{00}=-{{1}\over{2}}\partial_t h_{00}=0 \ \ \ and \ \ \ K^0_{00}=2g^{0k}T^0_{k0}g_{00}=0 \eqno{(36)}$$ while $${(\Gamma_{LC})}^i_{00}=\partial_i\phi \ \ \ with \ \ \  \phi={{1}\over{2}}h_{00} \ \ \ and \ \ \ K^i_{00}=2(\eta^{ik}-h^{ik})T^0_{k0}.\eqno{(37)}$$ Then $${{d^2\vec{x}}\over{dt^2}}=-\nabla\phi+{\bf H}\vec{T} \eqno{(38)}$$ with $${\bf H}=\pmatrix{1 & h^{12} & h^{13} \cr h^{12} & 1 & h^{23} \cr h^{13} & h^{23} & 1} \ \ \ and \ \ \ \vec{T}=\pmatrix{T^0_{10} \cr T^0_{20} \cr T^0_{30}}.\eqno{(39)}$$ 

\

i) If the three ${T^0_{k0}}$'s are of the same order of magnitude, $${{d^2\vec{x}}\over{dt^2}}=-\nabla\phi+2I\vec{T} \eqno{(40)}$$ ($I$ the identity matrix) and then $\vec{T}=\vec{0}$. 

\

ii) If one does not assume i) but nevertheless imposes the Newtonian limit, then $${\bf H}\vec{T}=0 \eqno{(41)}$$ which implies, for a non-vanishing $\vec{T}$, $det({\bf H})=1+O({h_{\mu\nu}}^2)=0$, which is impossible; therefore one has $\vec{T}=0$ again. 

\

In summary, the validity of the Newtonian limit implies $$T^0_{k0}=0, \ \ \ k=1,2,3 \eqno{(42)}$$ which is compatible with a totally antisymmetric $T_{\mu\nu\sigma}$. The Poisson equation for $\phi$ is consistent with these values of the ${T^0_{k0}}$'s [11].

\

{\bf 5. Final comments}

\

Upper bounds for the absolute values of components of the Cartan torsion tensor, in particular for its totally antisymmetric part (axial components), have been reported in [12], where experimental searches for Lorentz violation in the solar system were interpreted entirely in terms of the presence of torsion. For the axial components minimally coupled to fermions, the obtained values are extremely small, of the order of $10^{-15}$ m$^{-1}\simeq 10^{-31}$ Gev. 

\

A similar result as ours concerning the validity of the equivalence principle in the presence of a totally antisymmetric torsion has been recently obtained in [13]; a more deep investigation on this tensor can be found in [14] by the same author.

\

{\bf Acknowledgements}

\

This work was partially supported by the project PAPIIT-IN118609, DGAPA-UNAM, M\'exico. The author thanks M. Leston for a challenging discussion.

\

{\bf References}

\

[1] Cartan E. Sur une g\'eneralisation de la notion de courbure de Riemann et les espaces $\grave{a}$ torsion. Comptes Rendus Acad. Sci. 1922; 174: 593-595.

\

[2] Hehl FW, von der Heyde P, Kerlick GD, Nester JM. General relativity with spin and torsion: Foundations and prospects. Rev. Mod. Phys. 1976; 48: 393-416.

\

[3] von der Heyde P. The Equivalence Principle in the $U_4$ Theory of Gravitation. Lett. Nuovo Cimento. 1975; 14: 250-252.

\

[4] Hartley D. Normal frames for non-Riemannian connections. Class. Quantum Grav. 1995; 12: L103-105.

\

[5] Iliev BZ. Normal frames and the validity of the equivalence principle: I. Cases in a neighbourhood and at a point. J. Phys. A: Math. Gen. 1996; 29: 6895-6901.

\

[6] Nieto JA, Saucedo J, Villanueva VM. Relativistic top deviation equation and gravitational waves. Phys. Lett. A. 2003; 312: 175-186.

\

[7] Garcia de Andrade LC. Nongeodesic motion of spinless particles in the teleparallel gravitational wave background. 2002; arXiv: gr-qc/0205120.

\

[8] Landau LD, Lifshitz EM. The Classical Theory of Fields. 4th ed. Amsterdam: Elsevier; 1975: 259.  

\

[9] Carroll S. Spacetime and Geometry. An introduction to General Relativity. San Francisco: Addison Wesley; 2004: 106-108.

\

[10] Hehl FW. How does one measure torsion of space-time? Phys. Lett. A. 1971; 36: 225-226.  

\

[11] Pauli W. Theory of Relativity. New York: Dover; 1981: 163.

\

[12] Kosteleck\'y VA, Russell N, Tasson JD. Constraints on Torsion from Bounds on Lorentz Violation. Phys. Rev. Lett. 2008; 100: 111102.

\

[13] Fabbri L. On the Principle of Equivalence. 2009; arXiv: gr-qc/0905.2541.

\
 
[14] Fabbri L. On a Completely Antisymmetric Cartan Torsion Tensor. Annales de la Fondation Louis de Broglie. 2007; 32: 215-228; arXiv: gr-qc/0608090.

\

\

\

\

\

\

\

\

\

\

\

\

\

$(*)$ e-mail: socolovs@nucleares.unam.mx

\

\end